# The Research Productivity of Small Telescopes and Space Telescopes

F. A. Ringwald and John M. Culver

Department of Physics

California State University, Fresno

2345 E. San Ramon Ave., M/S MH37

Fresno, CA 94740-8031

E-mail: ringwald@csufresno.edu

Rebecca L. Lovell, Sarah Abbey Kays, and Yolanda V. Torres

Department of Physics and Space Sciences

Florida Institute of Technology

150 West University Boulevard

Melbourne, FL 32901

**ABSTRACT**. We present statistics on the research productivity of astronomical telescopes. These were compiled by finding papers in which new data were presented, noting which telescopes were used, and then counting the number of papers, number of pages, and other statistics. The journals used were the Astronomical Journal, the Astrophysical Journal (including the Letters and Supplements), and the Publications of the Astronomical Society of the Pacific. We also compiled citations from the Science Citation Index.

This work was designed to be similar to that of Trimble (1995), except that more recent journals (from 1995) and citations (from 1998) were used. We also did not restrict our sample to large telescopes only: we included all telescopes from which new data were presented, the smallest of which was a 0.1-m. The data were gathered by first-year work-study undergraduates, who were instructed to include data for all telescopes for which they found new data were included in the journals. A by-product of this research was therefore the relative productivity of ground-based versus space telescopes, and the relative productivity of radio and other telescopes across the spectrum, versus optical telescopes.

**Introduction**

Many small telescopes, with apertures of less than 3 m, are being closed at U. S. national and other observatories (e.g. ESO, NOAO). (See Table 1.) This is disturbing, since small telescopes are still capable of first-class science, often at far less cost than for large telescopes. For example, the MACHO, EROS, and OGLE gravitational microlensing projects all use 1.3-m or smaller telescopes (Alcock et al. 1995; Beaulieu et al. 1995; Paczynski et al. 1994). Mayor and Queloz (1995) used the 1.9-m telescope at Haute Provence for the first extrasolar planet detection, for a normal star; Marcy and Butler (1996) used the Lick 3-m and the 0.6-m coudé feed at Lick. Multi-longitude stellar seismology campaigns (Nather et al. 1990) mainly use telescopes in "the lowly 1-m class" (as in the words of R. E. Nather). The Center for Backyard Astrophysics is a worldwide network mainly composed of 10-inch amateur telescopes dedicated to photometry of cataclysmic variables, and has produced many results on accretion disk physics (e.g. Patterson 2001). A single automated 0.4-m telescope called "RoboScope," in Indiana, a state not usually noted as an astronomical site, has been used to discover several unexpected accretion disk phenomena (Robertson, Honeycutt, & Turner 1995; Honeycutt, Robertson, & Turner 1998). Other examples of competitive small-telescope science include echo mapping of active galactic nuclei (Welsh & Horne 1991) and finding stellar-mass black holes (Shahbaz et al. 1994).

Small telescopes can hold their own with larger instruments since more time is available on them. This makes possible monitoring campaigns, areal surveys, and time-resolved campaigns, particularly if the telescopes are networked or automated—all difficult to carry out with larger telescopes, for which even small amounts of telescope time are in great demand.

**Data on Telescope Productivity**

How productive are small telescopes? Three first-year undergraduates (RLL, SAK, YVT) spent a school year looking at papers in the Astronomical Journal, the Astrophysical Journal (including the main journal and the Letters), and the Publications of the Astronomical Society of the Pacific. A fourth, more advanced student (JMC) spent half a semester compiling similar data for the ApJ Supplements. These were the same journals used by Trimble (1995) in her study of the productivity of large, American optical telescopes.

Following Trimble (1995), who studied telescope productivity in 1990-91, and Abt (1985), who did so for 1980-81, the students read every paper that appeared in 1995 well enough to answer the following questions: Were new data presented? Which telescopes were used? How many pages did each paper have? How many citations did each paper have? They also counted citations from the Science Citation Index for 1998, the most recent complete year available when the project was begun in 1999 Fall.

Identifying individual telescopes was quickly found to be a problem, because these were first-year students, who had not yet heard of many telescopes. The PI (FAR) therefore advised them to collect data for all telescopes. This included several types of instruments not considered by Trimble (1995): (1) Large and small telescopes (Trimble included only large [> 2 m] telescopes); (2) Space telescopes (Trimble included only ground-based); (3) Instruments operating at radio and other wavelengths (Trimble included only optical/near-infrared). Therefore, as a by-product of examining small telescopes, we compiled productivity statistics for space-based and other telescopes.

American journals were chosen, to reflect American telescopes. These statistics are not necessarily reliable for other telescopes: for example, papers from British and Australian telescopes are often published in the *Monthly Notices of the Royal Astronomical Society*, and papers from European telescopes are often published in *Astronomy & Astrophysics*. Solar telescopes were included, but are probably under-represented, as the journals *Solar Physics*, *Journal of Geophysical Research*, and *Icarus* were not used. Solar telescopes have an advantage, however, when correlations with collecting area are considered: the Sun is so bright, apertures often need not be large. Radio telescopes and interferometers were included, but have a disadvantage for any statistic involving telescope collecting area, since they are so large. They were not included in comparisons involving area anyway, because it is often unclear how to make fair comparisons involving interferometers.

For each telescope, the total number of papers was counted, listed in Table 2 as "total papers." Table 3 lists similar data, but only for ground-based, optical/near-IR telescopes. Since many papers include new data taken from many telescopes, following Trimble (1995), we computed the weighted number of papers, listed in Tables 2 and 3 as "papers (weighted)," by assigning equal weight to each telescope (large and small) used in each project. This weighted number of papers means that if three telescopes were used in one paper, each telescope is credited with 1/3 of a paper, for that paper. Whether this is a fair comparison between telescopes is an open question. Giving equal weights may over-count or under-count the relative importance of small telescopes. Some research would not be possible without large telescopes, e.g. high-resolution spectroscopy of faint objects. Other research would never be carried out without small telescopes, e.g. surveys or time-resolved campaigns. Since we wanted our results to be comparable to those of Trimble (1995) and Abt (1985), we retained this convention.

The statistic $f$, included in Tables 2 and 3, is papers (weighted) divided by the total number of papers. It shows how often a given telescope is used in collaborative projects, involving other telescopes. If $f = 1$, the telescope is always used only by itself. If $f \ll 1$, it is often used in large projects involving many telescopes. Figure 1 shows $f$ plotted against aperture for all ground-based, optical/near-infrared telescopes. Figure 2 shows the same, for small telescopes only. There is one outlying point from the 10-m Keck telescope, which has the highest value of this statistic for any instrument, $f = 0.81$. This might have been expected, for a uniquely large and (in 1995) relatively new instrument: much commissioning science, involving only the unique capabilities of that instrument, was being done at the time. If one excludes the point from Keck, both Figures 1 and 2 show only a slight correlation ($R < 0.1$ for a linear regression) between $f$ and aperture (or area) for ground-based telescopes, so perhaps it is fair to give equal weights to the contributions of different telescopes. Large (4-m) and medium-size (2-3-m) telescopes were most often used with other telescopes, with $0.5 < f < 0.6$, giving the distribution a triangular shape. Small (0.3–1-m) telescopes have a nearly uniform distribution in $f$, with the largest spread in $f$ of any kind of instruments, showing their versatility: they are used both by themselves ($f = 0.76$ for the Mt. Laguna 1.0-m) and in combination with other telescopes ($f = 0.43$ for the KPNO 0.9-m). The clustering of points in the bottom-right corner of Figure 2 shows that the very smallest telescopes (0.1–0.4-m) are usually, but not always, used in large campaigns. There were only six campaigns involving ten or more telescopes: one was a VLBI radio campaign (Xu et al. 1995), two were multiwavelength campaigns on AGNs (Courvoisier et al. 1995; McDowell et al. 1995), one was a Whole Earth Telescope, multi-longitude seismological campaign (Kawaler et al. 1995), and two were done primarily with small university and amateur telescopes, organized by professionals (Hall et al. 1995; Kaye et al. 1995).

Statistics involving areas of the telescopes were also compiled (see Tables 2, 3, 12, and 13). This is of interest since Abt (1980) noted that telescope costs often correlate with collecting area.

**Results**

The journals under review and published in 1995 contained new data from 292 telescopes of all kinds. The smallest was a 0.1-m (4-inch) at Dublin Observatory in Delaware (Hall et al. 1995). There were 2211 total papers, 1322 weighted papers, and 4088 total citations, for all telescopes.

In 1995, space instruments produced 354 weighted papers, 480 total papers, 3294 pages, and 1233 citations. We summarize the data from space instruments in Table 4. Since three of four Nobel Prizes given for observational astronomy were for radio observations (1974 to Hewish and Ryle; 1978 to Penzias and Wilson; 1993 to Hulse and Taylor; the 2002 prize went to Giacconi for X-ray astronomy, and to Davis and Koshiba for neutrino astronomy), we also summarize results from radio telescopes, including millimeter- and sub-millimeter wave, in Table 5. We did not subdivide wavebands further, however, because where they begin and end is often a matter of opinion, driven by detector technology, e.g., 7800-10,000 Ångströms was called near-infrared in the days of photography (Keenan & Hynek 1950), but now is often called visible light since CCDs can detect it, even though human eyes can't.

Table 3 lists telescopes with > 9.0 citations, following Trimble (1995), sorted by number of papers. The data from these 39 instruments are summarized in Table 6, along with those of Trimble (1995). Most of these data are similar, so that between 1990-91 (Trimble's sample) and 1995 (ours), use patterns between telescopes changed relatively little. The main difference was that the total number of papers published for our sample was 1.3 times larger. This is not difficult to understand: although Trimble's sample covered a time span 1.5 times longer than ours, it included data from only 16 telescopes.

Table 2 shows that *Hubble Space Telescope* is significantly more productive than any other instrument, with regard to numbers of papers, pages, and citations. It also does well with statistics normalized by telescope area (Table 12). Table 2 also shows that other expensive facilities, such as the VLA, *ROSAT*, and *CGRO* are also among the most productive facilities.

Tables 7 (number of pages) and 8 (number of citations) correlate with Table 2 (total and weighted number of papers), since these tables include many of the same instruments (34/38, or 90%, for Table 7, and 33/38, or 87%, for Table 8). That the numbers of papers, pages, and citations would all correlate with each other is not surprising: Abt (1984) found this too. So did science historian Derek de Solla Price, who showed that quantity and quality often go together in scientific publication of all kinds (Price 1972, 1986). In particular, Price showed that on average, scientists in all fields who are doing the most important work are most often the scientists doing the most work, and that the most productive scientists are also the scientists that most often produce the most important science. This has also been confirmed by Abt (2000) and by Burstein (2000).

A useful model for this, elaborated on by Price, is Alfred J. Lotka's Law of Scientific Productivity, which states that the number of scientists with just $n$ publications is proportional to $1/n^2$ (Lotka 1926; Price 1972; Price 1986, p. 38). Our data, plotted in Figure 3, show that this roughly applies to telescopes, too, although the curve is steeper, fitting a $1/n^{3.4 \pm 0.1}$ law. But then, as Price (1986) points out, the high scorers will be exceptions anyway, so that excluding the *HST*, VLA, and *ROSAT* data, our data fit a $1/n^{1.5 + 0.1}$ law.

Table 9 (pages/paper), Table 10 (citations/page) and Table 11 (citations/paper) all show a mix of instruments significantly different from Tables 2 and 7-8. This is probably due to these statistics being capable of being strongly affected by single, highly cited papers, from instruments with relatively few total papers. Still, Table 10 (citations/page) features well-known, highly cited work, e.g., the MACHO project with the 1.3-m at Mount Stromlo (Alcock et al. 1995), and the discovery of Trans-Neptunian Objects, with the U. Hawaii 2.2-m (Jewitt & Luu 1995).

Table 12 lists papers, pages, and citations, all divided by telescope collecting area. A small space telescope, the 0.45-m *International Ultraviolet Explorer*, leads this list. *Hubble Space Telescope* also places high on this list, reassuring because it is such an expensive facility. Still, Table 12 is dominated by smaller instruments, as is Table 13, which lists similar data for ground-based optical/near-infrared telescopes only. These smaller instruments are often automated (e.g., the 0.25-m APT on Mt. Hopkins, or Indiana University's 0.4-m RoboScope) or networked or both (e.g., telescopes participating in Center for Backyard Astrophysics [CBA] campaigns). Nationally run Schmidt telescopes, such as the Burrell Schmidt on Kitt Peak and the CTIO Curtis Schmidt, place well in Tables 12 and 13. So do small, nationally run telescopes, such as the CTIO 0.9-m and the KPNO 0.9-m. The KPNO 2.1-m, CTIO 1.5-m, Palomar 1.52-m, and Steward 2.3-m (which, although productive, is always behind the public KPNO 2.1-m on the same mountain) are among the few others common to Table 12 and Tables 2 and 7-8.

The original question that motivated Abt (1985) and Trimble (1995) to carry out their studies was whether public (CTIO 4-m, KPNO 4-m) or private (Palomar 5-m, Lick 3-m) large telescopes are more productive. Both Abt and Trimble found they were about even. Tables 2 and 7-8 show the public telescopes to be ahead, at least in the optical/near-infrared. In Figure 3, the CTIO 4-m (the Blanco telescope) and the KPNO 4-m (the Mayall telescope) are the most productive ground-based optical/near-infrared telescopes of any kind, with 33.4 and 32.1 weighted papers in 1995, versus 25.5 for the Hale 5-m, and 18.4 for the Lick Shane 3-m (and 27.8 for the Canada-France-Hawaii 3.6-m, remarkable since it is primarily a foreign telescope: one might have expected the Canadian observers to have published in the *MNRAS,* and the French observers to have published in *A&A*). Normalizing by telescope aperture still favors the public telescopes, with Table 2 showing the CTIO 4-m to have produced 2.7 papers/area, versus the KPNO 4-m at 2.6, the Hale 5-m at 1.3, and the Lick 3-m at 2.6, and CFHT at 3.6.

However, none of these five telescopes even place in Tables 12 or 13, which is dominated by much smaller instruments. Among the telescopes common to Tables 2 and 12 are the CTIO 0.9-m, the KPNO 0.9-m, the KPNO 2.1-m, and the Palomar 1.52-m. Cursory examination of Figure 3 might lead one to conclude that, because most results come from only a few telescopes, one might therefore get away with supporting only those few telescopes. That small, inexpensive telescopes can and do place near the top contradicts this. Furthermore, when normalized by area, which to a large degree reflects costs, small telescopes come out distinctly ahead: not one of the telescopes in Tables 12 or 13 has a 3-m or larger aperture.

We therefore conclude that it is a tragedy for science that many observatories are closing their small telescopes, and not replacing them. Even if they are often handed over to private institutions, these telescopes do not become as productive as they are when they are open to proposals selected primarily by scientific merit. This is especially so for instruments with replacements that are not immediately forthcoming.

**Acknowledgments**

We thank the Federal Work-Study Program at Florida Institute of Technology for financial support for the students, and an American Astronomical Society Small Research Grant.

**Figure 1. Ground-based, optical/near-IR telescopes**

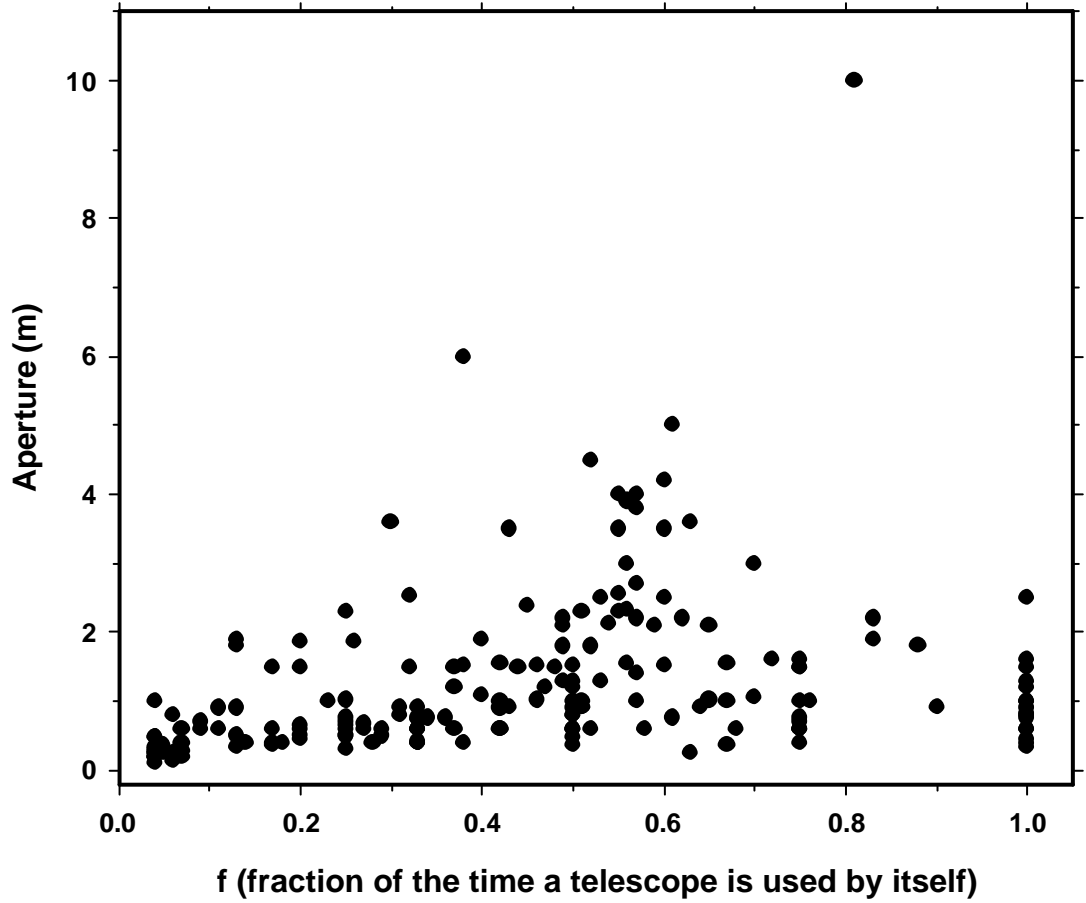

Figure 1. Ground-based, optical/near-IR telescopes.

# Figure 2. Small telescopes only

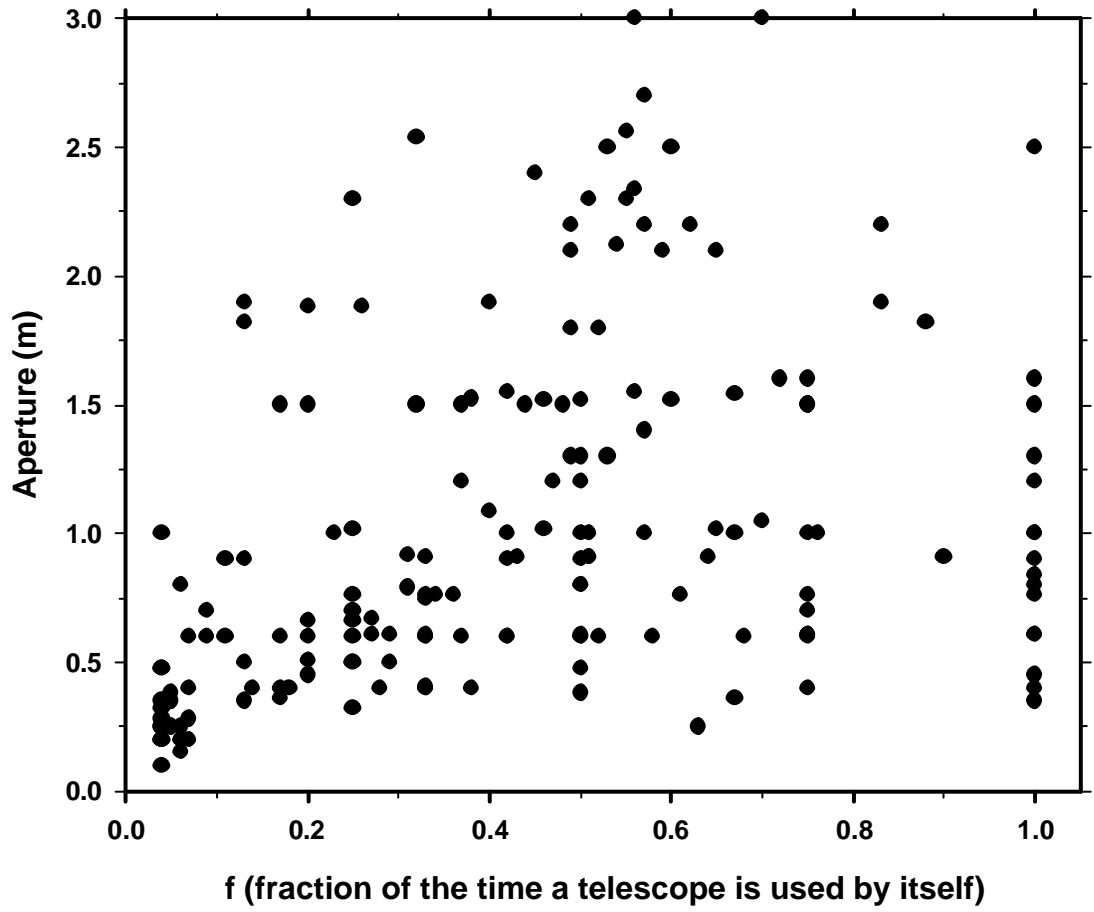

Figure 2. Same as Figure 1, but for small telescopes only.

# Figure 3. The research productivity of telescopes.

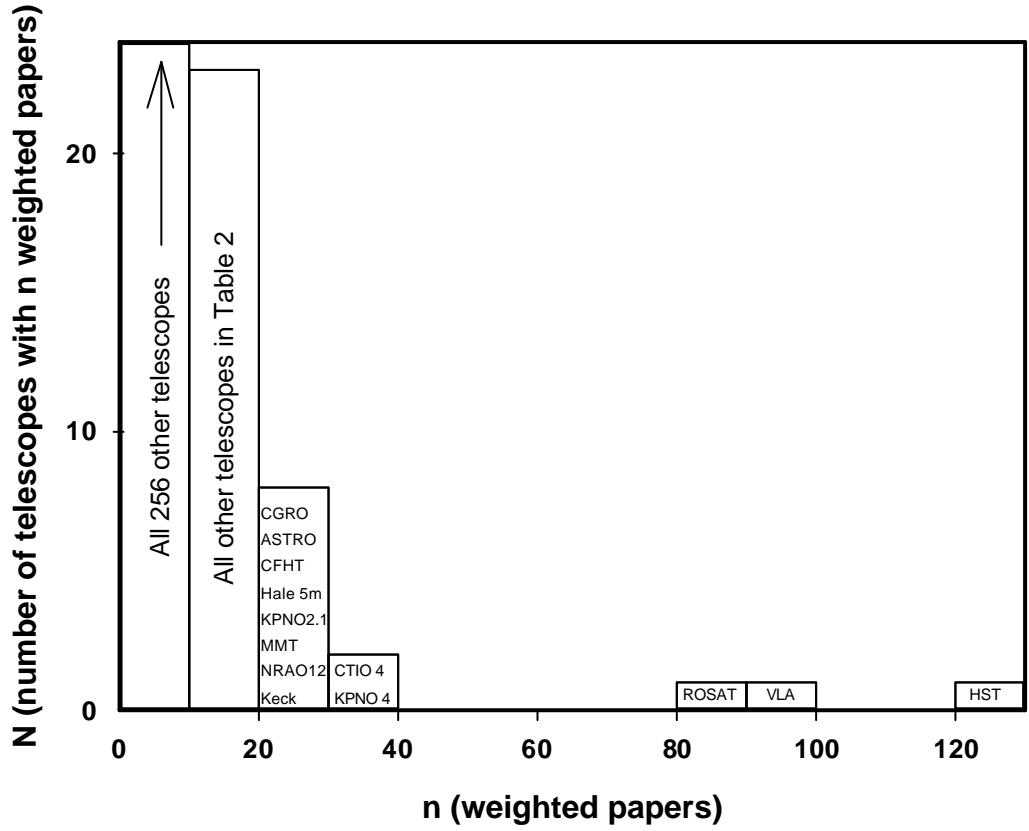

Figure 3. The research productivity of telescopes.

**Table 1. U. S. National Optical Astronomy Observatories (NOAO) telescopes, in 1995 and in 2002**

| NOAO in 1995 | NOAO in 2002 |
|---|---|
| N (Kitt Peak, Arizona) | N (Kitt Peak and elsewhere) |
| 4-m Mayall | 8-m Gemini N (50%) |
| 3.5-m WIYN (40%) | 4-m Mayall |
| 2.1-m | 3.5-m WIYN (40%) |
| 0.9-m coude feed | 9.2-m Hobby-Eberly Telescope (7%) |
| 0.9-m | 6.5-m MMT (7%) |
| 0.6-m Burrell Schmidt | 2.1-m |
| 1.3-m | 0.9-m (25%) |
| | |
| S (Cerro Tololo, Chile) | S (Cerro Tololo & Cerro Pachon) |
| 4-m Blanco | 8-m Gemini S (50%) |
| 1.5-m | 4-m Blanco |
| 0.9-m | 4-m SOAR (30%) (not yet online) |
| 1-m (50%) | 1.5-m (25% planned) |
| 0.6-m Curtis Schmidt | 1.3-m 2MASS (25 % planned) |
| 0.6-m Lowell | 1-m YALO (25% planned) |
| (0.4-m USNO) | 0.9-m (25% planned) |
| (0.4-m MPI) | |

**Table 2. Top telescopes, sorted by number of papers (weighted)**

| | total papers | f | papers (weighted) | pages | citations | pages/ paper | citations/ page | citations/ paper | papers/ area | pages/ area | citations/ area | |
|---|---|---|---|---|---|---|---|---|---|---|---|---|
| HST all instruments (2.4) | 163 | 0.79 | 129.1 | 1306 | 484.4 | 10.11 | 0.37 | 3.75 | 28.5 | 288.7 | 107.1 | space |
| VLA | 129 | 0.73 | 94.7 | 952 | 250.0 | 10.05 | 0.26 | 2.64 | | | | radio |
| ROSAT (all instruments) | 110 | 0.75 | 82.1 | 747 | 290.3 | 9.10 | 0.39 | 3.54 | | | | space |
| CTIO Blanco (4) | 61 | 0.55 | 33.4 | 373 | 108.9 | 11.17 | 0.29 | 3.26 | 2.7 | 29.7 | 8.7 | |
| KPNO Mayall (4) | 56 | 0.57 | 32.1 | 415 | 142.4 | 12.93 | 0.34 | 4.44 | 2.6 | 33.0 | 11.3 | |
| CGRO (all instruments) | 35 | 0.83 | 29.1 | 251 | 148.0 | 8.63 | 0.59 | 5.09 | | | | space |
| ASTRO (all instruments) | 34 | 0.84 | 28.6 | 194 | 56.5 | 6.79 | 0.29 | 1.98 | | | | space |
| CFHT (3.6) | 44 | 0.63 | 27.8 | 294 | 101.9 | 10.57 | 0.35 | 3.66 | 2.7 | 28.9 | 10.0 | |
| Palomar Hale (5) | 42 | 0.61 | 25.5 | 386 | 95.3 | 15.11 | 0.25 | 3.73 | 1.3 | 19.6 | 4.9 | |
| KPNO (2.1) | 52 | 0.49 | 25.5 | 281 | 70.4 | 11.03 | 0.25 | 2.77 | 7.3 | 81.1 | 20.3 | |
| MMT (4.5) | 48 | 0.52 | 25.0 | 285 | 97.7 | 11.41 | 0.34 | 3.92 | 1.6 | 17.9 | 6.1 | |
| NRAO (12) mm-wave | 32 | 0.71 | 22.9 | 222 | 34.7 | 9.70 | 0.16 | 1.52 | 0.2 | 2.0 | 0.3 | mm |
| Keck I & II (10) | 27 | 0.81 | 22.0 | 125 | 106.3 | 5.71 | 0.85 | 4.84 | 0.3 | 1.6 | 1.4 | |
| Steward (2.3) | 37 | 0.51 | 18.9 | 190 | 55.2 | 10.07 | 0.29 | 2.92 | 4.5 | 45.8 | 13.3 | |
| IUE (0.45) | 33 | 0.57 | 18.9 | 227 | 62.7 | 12.01 | 0.28 | 3.32 | 118.7 | 1426.1 | 394.2 | space |
| Lick Shane (3) | 33 | 0.56 | 18.4 | 213 | 57.1 | 11.58 | 0.27 | 3.11 | 2.6 | 30.1 | 8.1 | |
| Palomar (1.52) | 27 | 0.60 | 16.1 | 211 | 44.3 | 13.11 | 0.21 | 2.75 | 8.9 | 116.2 | 24.4 | |
| IRTF (3) | 22 | 0.70 | 15.5 | 123 | 43.0 | 7.95 | 0.35 | 2.77 | 2.2 | 17.4 | 6.1 | |
| AAT (3.9) | 27 | 0.56 | 15.0 | 142 | 51.6 | 9.47 | 0.36 | 3.44 | 1.3 | 11.9 | 4.3 | |
| Arecibo (305) | 17 | 0.87 | 14.8 | 154 | 35.3 | 10.39 | 0.23 | 2.38 | 0.0 | 0.0 | 0.0 | radio |
| JCMT (15) sub-mm | 21 | 0.68 | 14.3 | 190 | 28.2 | 13.32 | 0.15 | 1.98 | 0.1 | 1.1 | 0.2 | sub-mm |
| CTIO (1.5) | 27 | 0.48 | 12.9 | 151 | 38.1 | 11.72 | 0.25 | 2.96 | 7.3 | 85.4 | 21.5 | |
| CTIO (0.9) | 25 | 0.51 | 12.8 | 152 | 25.4 | 11.94 | 0.17 | 1.99 | 19.6 | 234.2 | 39.0 | |
| WHT (4.2) | 21 | 0.60 | 12.6 | 101 | 31.0 | 8.01 | 0.31 | 2.46 | 0.9 | 7.3 | 2.2 | |
| Caltech Sub-mm Obs (10) sub-mm | 19 | 0.65 | 12.3 | 139 | 27.5 | 11.23 | 0.20 | 2.23 | 0.2 | 1.8 | 0.4 | sub-mm |
| BIMA | 16 | 0.77 | 12.3 | 86 | 26.3 | 7.02 | 0.31 | 2.14 | | | | mm |
| Yohkoh solar | 15 | 0.78 | 11.8 | 109 | 32.0 | 9.26 | 0.29 | 2.72 | | | | solar |
| UKIRT (3.8) | 20 | 0.57 | 11.5 | 104 | 57.0 | 9.05 | 0.55 | 4.96 | 1.0 | 9.2 | 5.0 | |
| INT (2.5) | 19 | 0.60 | 11.5 | 123 | 32.5 | 10.76 | 0.26 | 2.84 | 2.3 | 25.1 | 6.6 | |
| KPNO (0.9) | 25 | 0.43 | 10.8 | 112 | 36.0 | 10.33 | 0.32 | 3.33 | 16.6 | 171.9 | 55.4 | |
| IRAM (30) mm-wave | 18 | 0.59 | 10.7 | 84 | 46.9 | 7.81 | 0.56 | 4.38 | 0.0 | 0.1 | 0.1 | mm |
| VLBA | 19 | 0.56 | 10.6 | 110 | 38.1 | 10.37 | 0.35 | 3.61 | | | | radio |
| NRAO (43) radio | 18 | 0.59 | 10.5 | 151 | 29.6 | 14.33 | 0.20 | 2.80 | 0.0 | 0.1 | 0.0 | radio |
| KPNO (1.3) | 21 | 0.49 | 10.3 | 116 | 26.0 | 11.32 | 0.22 | 2.53 | 7.7 | 87.5 | 19.6 | |
| IRAS | 16 | 0.64 | 10.3 | 91 | 30.4 | 8.89 | 0.33 | 2.97 | | | | space |
| Las Campanas Dupont (2.5) | 19 | 0.53 | 10.0 | 169 | 39.1 | 16.86 | 0.23 | 3.91 | 2.0 | 34.3 | 8.0 | |
| MDMO Hiltner (2.4) | 21 | 0.45 | 9.6 | 100 | 21.4 | 10.46 | 0.21 | 2.24 | 2.1 | 22.1 | 4.7 | |
| CTIO Yale (1.02) | 20 | 0.46 | 9.3 | 107 | 16.2 | 11.53 | 0.15 | 1.75 | 11.3 | 130.7 | 19.9 | |
| Kuiper Airborne (KAO) (0.91) | 10 | 0.90 | 9.0 | 85 | 25.5 | 9.39 | 0.30 | 2.83 | 13.8 | 129.9 | 39.2 | aircraft |
| McDonald (2.7) | 15 | 0.57 | 8.5 | 124 | 37.6 | 14.50 | 0.30 | 4.40 | 1.5 | 21.6 | 6.6 | |
| Other | 797 | 0.48 | 405 | 4141 | 1107 | 10.22 | 0.27 | 2.73 | | | | |
| Total (or average) | 2211 | 0.50 | 1322 | 13632 | 4088 | 10.31 | 0.30 | 3.09 | | | | |

**Table 3. Top ground-based, optical/IR telescopes, sorted by number of papers (weighted)**

| | total papers | f | papers (weighted) | pages | citations | pages/ paper | citations/ page | citations/ paper | papers/ area | pages/ area | citations/ area |
|---|---|---|---|---|---|---|---|---|---|---|---|
| CTIO Blanco (4) | 61 | 0.55 | 33.4 | 373 | 108.9 | 11.17 | 0.29 | 3.26 | 2.7 | 29.7 | 8.7 |
| KPNO Mayall (4) | 56 | 0.57 | 32.1 | 415 | 142.4 | 12.93 | 0.34 | 4.44 | 2.6 | 33.0 | 11.3 |
| CFHT (3.6) | 44 | 0.63 | 27.8 | 294 | 101.9 | 10.57 | 0.35 | 3.66 | 2.7 | 28.9 | 10.0 |
| Palomar Hale (5) | 42 | 0.61 | 25.5 | 386 | 95.3 | 15.11 | 0.25 | 3.73 | 1.3 | 19.6 | 4.9 |
| KPNO (2.1) | 52 | 0.49 | 25.5 | 281 | 70.4 | 11.03 | 0.25 | 2.77 | 7.3 | 81.1 | 20.3 |
| MMT (4.5) | 48 | 0.52 | 25.0 | 285 | 97.7 | 11.41 | 0.34 | 3.92 | 1.6 | 17.9 | 6.1 |
| Keck I & II (10) | 27 | 0.81 | 22.0 | 125 | 106.3 | 5.71 | 0.85 | 4.84 | 0.3 | 1.6 | 1.4 |
| Steward (2.3) | 37 | 0.51 | 18.9 | 190 | 55.2 | 10.07 | 0.29 | 2.92 | 4.5 | 45.8 | 13.3 |
| Lick Shane (3) | 33 | 0.56 | 18.4 | 213 | 57.1 | 11.58 | 0.27 | 3.11 | 2.6 | 30.1 | 8.1 |
| Palomar (1.52) | 27 | 0.60 | 16.1 | 211 | 44.3 | 13.11 | 0.21 | 2.75 | 8.9 | 116.2 | 24.4 |
| IRTF (3) | 22 | 0.70 | 15.5 | 123 | 43.0 | 7.95 | 0.35 | 2.77 | 2.2 | 17.4 | 6.1 |
| AAT (3.9) | 27 | 0.56 | 15.0 | 142 | 51.6 | 9.47 | 0.36 | 3.44 | 1.3 | 11.9 | 4.3 |
| CTIO (1.5) | 27 | 0.48 | 12.9 | 151 | 38.1 | 11.72 | 0.25 | 2.96 | 7.3 | 85.4 | 21.5 |
| CTIO (0.9) | 25 | 0.51 | 12.8 | 152 | 25.4 | 11.94 | 0.17 | 1.99 | 19.6 | 234.2 | 39.0 |
| WHT (4.2) | 21 | 0.60 | 12.6 | 101 | 31.0 | 8.01 | 0.31 | 2.46 | 0.9 | 7.3 | 2.2 |
| UKIRT (3.8) | 20 | 0.57 | 11.5 | 104 | 57.0 | 9.05 | 0.55 | 4.96 | 1.0 | 9.2 | 5.0 |
| INT (2.5) | 19 | 0.60 | 11.5 | 123 | 32.5 | 10.76 | 0.26 | 2.84 | 2.3 | 25.1 | 6.6 |
| KPNO (0.9) | 25 | 0.43 | 10.8 | 112 | 36.0 | 10.33 | 0.32 | 3.33 | 16.6 | 171.9 | 55.4 |
| KPNO (1.3) | 21 | 0.49 | 10.3 | 116 | 26.0 | 11.32 | 0.22 | 2.53 | 7.7 | 87.5 | 19.6 |
| Las Campanas Dupont (2.5) | 19 | 0.53 | 10.0 | 169 | 39.1 | 16.86 | 0.23 | 3.91 | 2.0 | 34.3 | 8.0 |
| MDMO Hiltner (2.4) | 21 | 0.45 | 9.6 | 100 | 21.4 | 10.46 | 0.21 | 2.24 | 2.1 | 22.1 | 4.7 |
| CTIO Yale (1.02) | 20 | 0.46 | 9.3 | 107 | 16.2 | 11.53 | 0.15 | 1.75 | 11.3 | 130.7 | 19.9 |
| McDonald (2.7) | 15 | 0.57 | 8.5 | 124 | 37.6 | 14.50 | 0.30 | 4.40 | 1.5 | 21.6 | 6.6 |
| ESO NTT (3.5) | 14 | 0.60 | 8.5 | 64 | 18.0 | 7.55 | 0.28 | 2.13 | 0.9 | 6.6 | 1.9 |
| Las Campanas Swope (1.02) | 13 | 0.65 | 8.4 | 95 | 25.8 | 11.33 | 0.27 | 3.06 | 10.3 | 116.8 | 31.5 |
| Lowell Perkins (1.8) | 15 | 0.52 | 7.8 | 97 | 19.7 | 12.49 | 0.20 | 2.54 | 3.1 | 38.1 | 7.7 |
| KPNO coude feed (0.9) | 12 | 0.64 | 7.7 | 130 | 18.0 | 16.93 | 0.14 | 2.34 | 11.8 | 200.5 | 27.7 |
| Mt. Laguna (1.0) | 10 | 0.76 | 7.6 | 54 | 9.9 | 7.16 | 0.18 | 1.31 | 9.7 | 69.2 | 12.6 |
| UHawaii (2.2) | 12 | 0.62 | 7.4 | 95 | 58.2 | 12.80 | 0.61 | 7.84 | 2.0 | 25.0 | 15.3 |
| McDonald (2.1) | 12 | 0.59 | 7.0 | 89 | 12.8 | 12.70 | 0.14 | 1.82 | 2.0 | 25.8 | 3.7 |
| McGraw-Hill (1.3) | 13 | 0.53 | 6.9 | 54 | 11.3 | 7.90 | 0.21 | 1.64 | 5.2 | 40.8 | 8.5 |
| Burrell Schmidt (0.6) | 10 | 0.63 | 6.3 | 55 | 9.3 | 8.67 | 0.17 | 1.47 | 22.4 | 194.2 | 33.0 |
| ESO (2.2) | 10 | 0.57 | 5.7 | 53 | 13.2 | 9.24 | 0.25 | 2.32 | 1.5 | 13.9 | 3.5 |
| Whipple (1.2) | 12 | 0.47 | 5.6 | 52 | 12.0 | 9.28 | 0.23 | 2.15 | 4.9 | 45.8 | 10.6 |
| Siding Spring (2.3) | 10 | 0.55 | 5.5 | 63 | 29.0 | 11.46 | 0.46 | 5.24 | 1.3 | 15.3 | 7.0 |
| Kiso Schmidt (1.05) | 7 | 0.70 | 4.9 | 54 | 10.0 | 10.94 | 0.19 | 2.04 | 5.7 | 61.9 | 11.5 |
| McMath-Pierce (2.1) | 7 | 0.65 | 4.6 | 60 | 17.0 | 13.16 | 0.28 | 3.71 | 1.3 | 17.4 | 4.9 |
| DAO (1.8) | 9 | 0.49 | 4.4 | 55 | 7.3 | 12.36 | 0.13 | 1.66 | 1.7 | 21.4 | 2.9 |
| Whipple (1.5) | 10 | 0.44 | 4.4 | 50 | 19.0 | 11.35 | 0.38 | 4.34 | 2.5 | 28.1 | 10.8 |
| Other | 390 | 0.44 | 175 | 1619 | 391 | 9.25 | 0.24 | 2.24 | | | |
| Total (or average) | 1275 | 0.46 | 672 | 7135 | 2017 | 10.61 | 0.28 | 3.00 | | | |

**Table 4. Space Instruments, sorted by number of papers (weighted)**

| | total papers | f | papers (weighted) | pages | citations | pages/ paper | citations/ page | citations/ paper |
|---|---|---|---|---|---|---|---|---|
| HST all instruments (2.4) | 163 | 0.79 | 129.1 | 1306 | 484.4 | 10.11 | 0.37 | 3.75 |
| ROSAT (all instruments) | 110 | 0.75 | 82.1 | 747 | 290.3 | 9.10 | 0.39 | 3.54 |
| CGRO (all instruments) | 35 | 0.83 | 29.1 | 251 | 148.0 | 8.63 | 0.59 | 5.09 |
| ASTRO (all instruments) | 34 | 0.84 | 28.6 | 194 | 56.5 | 6.79 | 0.29 | 1.98 |
| IUE (0.45) | 33 | 0.57 | 18.9 | 227 | 62.7 | 12.01 | 0.28 | 3.32 |
| Yohkoh solar | 15 | 0.78 | 11.8 | 109 | 32.0 | 9.26 | 0.29 | 2.72 |
| IRAS | 16 | 0.64 | 10.3 | 91 | 30.4 | 8.89 | 0.33 | 2.97 |
| ASCA | 8 | 0.81 | 6.5 | 44 | 36.5 | 6.77 | 0.83 | 5.62 |
| EUVE (all instruments) | 8 | 0.67 | 5.4 | 51 | 17.1 | 9.37 | 0.34 | 3.18 |
| EXOSAT (all instruments) | 11 | 0.36 | 4.0 | 37 | 9.2 | 9.35 | 0.25 | 2.32 |
| SIGMA/GRANAT | 5 | 0.73 | 3.7 | 26 | 7.7 | 7.09 | 0.29 | 2.09 |
| SAMPEX | 4 | 0.78 | 3.1 | 20 | 5.0 | 6.28 | 0.25 | 1.60 |
| Ginga | 3 | 0.83 | 2.5 | 20 | 2.5 | 7.80 | 0.13 | 1.00 |
| Atlas-1 (FAUST) | 3 | 0.83 | 2.5 | 22 | 6.0 | 8.80 | 0.27 | 2.40 |
| ORFEUS (1) | 4 | 0.58 | 2.3 | 22 | 4.8 | 9.50 | 0.22 | 2.07 |
| HEAO A-1 | 4 | 0.46 | 1.8 | 24 | 4.7 | 12.82 | 0.20 | 2.55 |
| Einstein (all instruments) | 5 | 0.26 | 1.3 | 4 | 0.9 | 3.32 | 0.22 | 0.73 |
| Spacelab 2 solar | 1 | 1.00 | 1.0 | 9 | 1.0 | 9.00 | 0.11 | 1.00 |
| NASA Sounding rocket | 1 | 1.00 | 1.0 | 7 | 0.0 | 7.00 | 0.00 | 0.00 |
| ONR-604 | 1 | 1.00 | 1.0 | 11 | 0.0 | 11.00 | 0.00 | 0.00 |
| Voyager 1 | 1 | 1.00 | 1.0 | 3 | 0.0 | 3.00 | 0.00 | 0.00 |
| COBE (all instruments) | 1 | 1.00 | 1.0 | 12 | 17.0 | 12.00 | 1.42 | 17.00 |
| Solar Max (all instruments) | 1 | 1.00 | 1.0 | 11 | 0.0 | 11.00 | 0.00 | 0.00 |
| UARS all instruments | 1 | 1.00 | 1.0 | 4 | 5.0 | 4.00 | 1.25 | 5.00 |
| STS-39 Far-UV Cameras | 1 | 1.00 | 1.0 | 21 | 1.0 | 21.00 | 0.05 | 1.00 |
| GOES | 1 | 0.50 | 0.5 | 8 | 0.5 | 16.00 | 0.06 | 1.00 |
| Spartan 201 (solar) | 1 | 0.50 | 0.5 | 2 | 6.0 | 4.00 | 3.00 | 12.00 |
| Voyager 2 | 1 | 0.25 | 0.3 | 1 | 0.0 | 5.00 | 0.00 | 0.00 |
| Uhuru | 1 | 0.17 | 0.2 | 2 | 0.3 | 9.00 | 0.22 | 2.00 |
| Ariel V | 1 | 0.17 | 0.2 | 2 | 0.3 | 9.00 | 0.22 | 2.00 |
| Salyut 6 | 1 | 0.13 | 0.1 | 1 | 0.0 | 5.00 | 0.00 | 0.00 |
| S81-1 | 1 | 0.13 | 0.1 | 1 | 0.0 | 5.00 | 0.00 | 0.00 |
| Spacelab 1 | 1 | 0.13 | 0.1 | 1 | 0.0 | 5.00 | 0.00 | 0.00 |
| Spacelab 3 | 1 | 0.13 | 0.1 | 1 | 0.0 | 5.00 | 0.00 | 0.00 |
| Cosmos 2022 | 1 | 0.13 | 0.1 | 1 | 0.0 | 5.00 | 0.00 | 0.00 |
| Total (or average) | 479 | 0.62 | 353 | 3288 | 1230 | 9.31 | 0.37 | 3.48 |

**Table 5. Top radio telescopes, sorted by number of papers (weighted)**

| | total papers | f | papers (weighted) | pages | citations | pages/ paper | citations/ page | citations/ paper | |
|---|---|---|---|---|---|---|---|---|---|
| VLA | 129 | 0.73 | 94.7 | 952 | 250.0 | 10.05 | 0.26 | 2.64 | radio |
| NRAO (12) mm-wave | 32 | 0.71 | 22.9 | 222 | 34.7 | 9.70 | 0.16 | 1.52 | mm |
| Arecibo (305) | 17 | 0.87 | 14.8 | 154 | 35.3 | 10.39 | 0.23 | 2.38 | radio |
| JCMT (15) sub-mm | 21 | 0.68 | 14.3 | 190 | 28.2 | 13.32 | 0.15 | 1.98 | sub-mm |
| Caltech Submillimeter Obs (10) | 19 | 0.65 | 12.3 | 139 | 27.5 | 11.23 | 0.20 | 2.23 | sub-mm |
| BIMA | 16 | 0.77 | 12.3 | 86 | 26.3 | 7.02 | 0.31 | 2.14 | mm |
| IRAM (30) mm-wave | 18 | 0.59 | 10.7 | 84 | 46.9 | 7.81 | 0.56 | 4.38 | mm |
| VLBA | 19 | 0.56 | 10.6 | 110 | 38.1 | 10.37 | 0.35 | 3.61 | radio |
| NRAO (43) radio | 18 | 0.59 | 10.5 | 151 | 29.6 | 14.33 | 0.20 | 2.80 | radio |
| Nobeyama (45) radio | 8 | 0.85 | 6.8 | 61 | 7.5 | 8.98 | 0.12 | 1.10 | radio |
| Five College (FCRAO) (14) radio | 11 | 0.57 | 6.3 | 133 | 27.7 | 21.11 | 0.21 | 4.38 | radio |
| Lovell (76) radio | 11 | 0.57 | 6.3 | 133 | 27.7 | 21.11 | 0.21 | 4.38 | radio |
| Parkes (64) radio | 11 | 0.57 | 6.3 | 133 | 27.7 | 21.11 | 0.21 | 4.38 | radio |
| Owens Valley mm array | 10 | 0.63 | 6.3 | 36 | 33.8 | 5.77 | 0.93 | 5.37 | mm |
| Nobeyama mm array (NMA) | 7 | 0.76 | 5.3 | 41 | 16.0 | 7.66 | 0.39 | 3.00 | mm |
| Haystack (37) radio | 10 | 0.49 | 4.9 | 30 | 13.1 | 6.05 | 0.44 | 2.67 | radio |
| ATCA | 6 | 0.78 | 4.7 | 38 | 14.0 | 8.21 | 0.37 | 3.00 | radio |
| Effelsberg (100) radio | 12 | 0.36 | 4.3 | 47 | 15.6 | 10.86 | 0.34 | 3.65 | radio |
| Australia Telescope (ATNF) | 6 | 0.48 | 2.9 | 25 | 7.5 | 8.60 | 0.30 | 2.60 | radio |
| Westerbork radio array | 8 | 0.27 | 2.1 | 27 | 9.7 | 12.79 | 0.36 | 4.55 | radio |
| Owens Valley (10.4) radio | 4 | 0.49 | 2.0 | 16 | 2.7 | 7.95 | 0.17 | 1.36 | radio |
| DRAO radio array | 3 | 0.48 | 1.5 | 13 | 3.3 | 8.97 | 0.25 | 2.28 | radio |
| Merlin array | 6 | 0.24 | 1.4 | 17 | 5.9 | 11.92 | 0.35 | 4.13 | radio |
| Molongo Obs Synth Telescope (MOST) | 2 | 0.67 | 1.3 | 14 | 2.0 | 10.50 | 0.14 | 1.50 | radio |
| Green Bank Interferometer radio | 2 | 0.67 | 1.3 | 12 | 5.3 | 8.75 | 0.46 | 4.00 | radio |
| ESO Swedish (15) sub-mm | 2 | 0.67 | 1.3 | 11 | 4.0 | 8.25 | 0.36 | 3.00 | sub-mm |
| RATAN RT-22 radio | 1 | 1.00 | 1.0 | 9 | 2.0 | 9.00 | 0.22 | 2.00 | radio |
| Nagoya (4) mm-wave | 1 | 1.00 | 1.0 | 5 | 0.0 | 5.00 | 0.00 | 0.00 | mm |
| JPL Goldstone radar | 1 | 1.00 | 1.0 | 4 | 1.0 | 4.00 | 0.25 | 1.00 | radar |
| Python (0.75) sub-mm | 1 | 1.00 | 1.0 | 4 | 3.0 | 4.00 | 0.75 | 3.00 | sub-mm |
| Hartbeesthoek (26) radio | 1 | 1.00 | 1.0 | 8 | 3.0 | 8.00 | 0.38 | 3.00 | radio |
| Tokyo-Nobeyama (0.6) sub-mm | 1 | 1.00 | 1.0 | 10 | 3.0 | 10.00 | 0.30 | 3.00 | sub-mm |
| Owens Valley (40) radio | 5 | 0.17 | 0.9 | 13 | 6.1 | 14.71 | 0.47 | 6.93 | radio |
| Deep Space Network (DSN) radio (70) | 2 | 0.35 | 0.7 | 5 | 5.6 | 7.14 | 1.12 | 8.00 | radio |
| Hat Creek (26) radio | 1 | 0.50 | 0.5 | 5 | 5.0 | 9.00 | 1.11 | 10.00 | radio |
| Inst Argentino de Radioastr (30) | 1 | 0.50 | 0.5 | 3 | 0.5 | 6.00 | 0.17 | 1.00 | radio |
| U Michigan radio (20) | 2 | 0.25 | 0.5 | 8 | 2.7 | 15.67 | 0.34 | 5.33 | radio |
| AT&T Bell Labs (7) radio | 2 | 0.20 | 0.4 | 6 | 3.2 | 15.50 | 0.52 | 8.00 | radio |
| Cambridge One-Mile radio | 1 | 0.33 | 0.3 | 9 | 1.3 | 27.00 | 0.15 | 4.00 | radio |
| Cambridge 5 km radio | 1 | 0.33 | 0.3 | 9 | 1.3 | 27.00 | 0.15 | 4.00 | radio |
| Algonquin radio | 1 | 0.33 | 0.3 | 7 | 0.7 | 21.00 | 0.10 | 2.00 | radio |
| Medicina radio (32) | 2 | 0.15 | 0.3 | 6 | 1.4 | 21.75 | 0.22 | 4.81 | radio |
| Noto radio (32) | 2 | 0.15 | 0.3 | 6 | 1.4 | 21.75 | 0.22 | 4.81 | radio |

| | | | | | | | | |
|---|---|---|---|---|---|---|---|---|
| Nancay radio | | 1 | 0.25 | 0.3 | 2 | 0.3 | 7.00 | 0.14 | 1.00 radio |
| Robledo radio (34) | | 1 | 0.20 | 0.2 | 2 | 0.4 | 8.00 | 0.25 | 2.00 radio |
| DSN (34) radio | | 1 | 0.20 | 0.2 | 3 | 2.6 | 15.00 | 0.87 | 13.00 radio |
| Total (or average) | | 437 | 0.56 | 280 | 2996 | 784 | 10.71 | 0.26 | 2.80 |

**Table 6. Summary of data from the 1995 AJ, ApJ**
**(including Letters and Supplements), and PASP**

| | All 292 telescopes of all kinds (Table 2) | Telescopes with > 9.0 citations (Table 2) (1) | From Trimble (1995) (2) | Space instruments (Table 5) (3) | Radio telescopes (Table 6) (4) |
|---|---|---|---|---|---|
| Total papers: | 2211 | 1399 | | 479 | 437 |
| Papers (weighted): | 1322 | 908 | 663 | 353 | 280 |
| Pages: | 13632 | 9367 | 7290 | 3288 | 2996 |
| Citations: | 4088 | 2943 | 2705 | 1230 | 784 |
| Mean pages per paper: | 10.31 | 10.30 | 11.00 | 9.31 | 10.71 |
| Mean citations per page: | 0.30 | 0.31 | 0.27 | 0.37 | 0.26 |
| Mean citations per paper: | 3.09 | 3.24 | 4.08 | 3.48 | 2.80 |

(1) These 39 telescopes include:
    11 large ground-based optical/IR telescopes (10-m to 3-m),
    11 small ground-based optical/IR telescopes (2.5-m to 0.9-m),
    9 ground-based radio (4), mm-wave (3), or sub-mm (2) telescopes,
    6 space instruments (*HST, ROSAT, CGRO, ASTRO, IUE, IRAS*),
    One solar instrument (the *Yohkoh* spacecraft),
    One other instrument (Kuiper Airborne Observatory).
(2) Trimble (1995) included results only from 16 large (> 2 m),
    optical/near-IR telescopes, all with > 9 citations in 1990-91.
(3) All cases are between 20-30% of the total for each.
(4) Including mm, sub-mm, and longer wavelengths

**Table 7. Top telescopes, sorted by number of pages**

|  | pages |  |
|---|---|---|
| HST all instruments (2.4) | 1306 | space |
| VLA | 952 | radio |
| ROSAT (all instruments) | 747 | space |
| KPNO Mayall (4) | 415 | |
| Palomar Hale (5) | 386 | |
| CTIO Blanco (4) | 373 | |
| CFHT (3.6) | 294 | |
| MMT (4.5) | 285 | |
| KPNO (2.1) | 281 | |
| CGRO (all instruments) | 251 | space |
| IUE (0.45) | 227 | space |
| NRAO (12) mm-wave | 222 | mm |
| Lick Shane (3) | 213 | |
| Palomar (1.52) | 211 | |
| ASTRO (all instruments) | 194 | space |
| Steward (2.3) | 190 | |
| JCMT (15) sub-mm | 190 | sub-mm |
| Las Campanas Dupont (2.5) | 169 | |
| Arecibo (305) | 154 | radio |
| CTIO (0.9) | 152 | |
| NRAO (43) radio | 151 | radio |
| CTIO (1.5) | 151 | |
| AAT (3.9) | 142 | |
| Caltech Submillimeter Obs (10) sub-mm | 139 | sub-mm |
| Five College (FCRAO) (14) radio | 133 | radio |
| Lovell (76) radio | 133 | radio |
| Parkes (64) radio | 133 | radio |
| KPNO coude feed (0.9) | 130 | |
| Keck I & II (10) | 125 | |
| McDonald (2.7) | 124 | |
| INT (2.5) | 123 | |
| IRTF (3) | 123 | |
| KPNO (1.3) | 116 | |
| KPNO (0.9) | 112 | |
| VLBA | 110 | radio |
| Yohkoh solar | 109 | solar |
| CTIO Yale (1.02) | 107 | |
| UKIRT (3.8) | 104 | |
| WHT (4.2) | 101 | |
| MDMO Hiltner (2.4) | 100 | |
| Other | 3956 | |
| Total | 13632 | |

**Table 8. Top telescopes, sorted by number of citations**

|  | citations |  |
|---|---|---|
| HST all instruments (2.4) | 484.4 | space |
| ROSAT (all instruments) | 290.3 | space |
| VLA | 250.0 | radio |
| CGRO (all instruments) | 148.0 | space |
| KPNO Mayall (4) | 142.4 |  |
| CTIO Blanco (4) | 108.9 |  |
| Keck I & II (10) | 106.3 |  |
| CFHT (3.6) | 101.9 |  |
| MMT (4.5) | 97.7 |  |
| Palomar Hale (5) | 95.3 |  |
| KPNO (2.1) | 70.4 |  |
| IUE (0.45) | 62.7 | space |
| UHawaii (2.2) | 58.2 |  |
| Lick Shane (3) | 57.1 |  |
| UKIRT (3.8) | 57.0 |  |
| ASTRO (all instruments) | 56.5 | space |
| Steward (2.3) | 55.2 |  |
| AAT (3.9) | 51.6 |  |
| IRAM (30) mm-wave | 46.9 | mm |
| Palomar (1.52) | 44.3 |  |
| IRTF (3) | 43.0 |  |
| Las Campanas Dupont (2.5) | 39.1 |  |
| VLBA | 38.1 | radio |
| CTIO (1.5) | 38.1 |  |
| McDonald (2.7) | 37.6 |  |
| ASCA | 36.5 | space |
| KPNO (0.9) | 36.0 |  |
| Arecibo (305) | 35.3 | radio |
| NRAO (12) mm-wave | 34.7 | mm |
| Owens Valley mm array | 33.8 | mm |
| INT (2.5) | 32.5 |  |
| Yohkoh solar | 32.0 | solar |
| WHT (4.2) | 31.0 |  |
| IRAS | 30.4 | space |
| NRAO (43) radio | 29.6 | radio |
| Siding Spring (2.3) | 29.0 |  |
| JCMT (15) sub-mm | 28.2 | sub-mm |
| Five College (FCRAO) (14) radio | 27.7 | radio |
| Lovell (76) radio | 27.7 | radio |
| Parkes (64) radio | 27.7 | radio |
| Other | 1035 |  |
| Total (or average) | 4088 |  |

**Table 9. Top telescopes, sorted by mean pages/paper**

|  | pages/paper |  |
|---|---|---|
| Onsala (26) | 52.0 |  |
| VU-TSU (0.4) | 37.8 |  |
| NRAO (91) radio | 36.2 | radio |
| Cambridge One-Mile radio | 27.0 | radio |
| Cambridge 5 km radio | 27.0 | radio |
| Palomar (0.45) Schmidt | 24.0 |  |
| UNM Capilla Peak (0.61) | 24.0 |  |
| MSU (0.6) | 24.0 |  |
| Okayama (1.88) | 23.0 |  |
| Medicina radio (32) | 21.8 | radio |
| Noto radio (32) | 21.8 | radio |
| Five College (FCRAO) (14) radio | 21.1 | radio |
| Lovell (76) radio | 21.1 | radio |
| Parkes (64) radio | 21.1 | radio |
| STS-39 Far-UV Cameras | 21.0 | space |
| James Gregory (0.9) | 21.0 |  |
| Algonquin radio | 21.0 | radio |
| Mt. Wilson (1.524) | 18.7 |  |
| MIRA (0.9) | 18.0 |  |
| Table Mt (0.5) | 18.0 |  |
| Cima Ekar (1.82) | 18.0 |  |
| Torun (0.9) | 18.0 |  |
| Lick CAT (0.6) | 17.6 |  |
| CASLEO (2.2) | 17.2 |  |
| Yale-Columbia (0.66) | 17.0 |  |
| Yale Southern (0.51) | 17.0 |  |
| KPNO coude feed (0.9) | 16.9 |  |
| Las Campanas Dupont (2.5) | 16.9 |  |
| Mt. Wilson (2.54) | 16.4 |  |
| GOES | 16.0 | space |
| Beijing A O (0.6) | 16.0 |  |
| Ege Univ. Obs. (0.48) | 16.0 |  |
| U Michigan radio (20) | 15.7 | radio |
| AT&T Bell Labs (7) radio | 15.5 | radio |
| UK Schmidt (1.2) | 15.2 |  |
| Pine Bluff (0.91) | 15.2 |  |
| Palomar Hale (5) | 15.1 |  |
| DSN (34) radio | 15.0 | radio |
| Owens Valley (40) radio | 14.7 | radio |
| McDonald (2.7) | 14.5 |  |
| Other | 8.4 |  |
| Total | 10.1 |  |

**Table 10. Top telescopes, sorted by citations/page**
This may or may not be a useful benchmark: note the small number of papers often involved.

| | citations/ page | papers (weighted) | pages | citations | |
|---|---|---|---|---|---|
| Spartan 201 (solar) | 3.00 | 0.5 | 2 | 6.0 | solar |
| CANGAROO (3.8) gamma ray | 2.75 | 0.5 | 2 | 5.5 | gamma-ray |
| COBE (all instruments) | 1.42 | 1.0 | 12 | 17.0 | space |
| UARS all instruments | 1.25 | 1.0 | 4 | 5.0 | space |
| Mt. Stromlo (1.3) | 1.19 | 2.0 | 16 | 19.0 | |
| Deep Space Network (DSN) radio (70) | 1.12 | 0.7 | 5 | 5.6 | radio |
| Hat Creek (26) radio | 1.11 | 0.5 | 5 | 5.0 | radio |
| Mees Solar coronagraph (0.25) | 1.06 | 0.8 | 6 | 6.3 | solar |
| Balloon-borne IR Carbon Explorer (BICE) | 1.00 | 1.0 | 4 | 4 | balloon |
| Teide (0.8) | 1.00 | 0.5 | 2 | 2.0 | |
| Owens Valley mm array | 0.93 | 6.3 | 36 | 33.8 | mm |
| DSN (34) radio | 0.87 | 0.2 | 3 | 2.6 | radio |
| Keck I & II (10) | 0.85 | 22.0 | 125 | 106.3 | |
| ASCA | 0.83 | 6.5 | 44 | 36.5 | space |
| Swedish solar VT (0.48) | 0.81 | 2.0 | 16 | 13.0 | |
| ESO Dutch (0.9) | 0.79 | 0.8 | 6 | 4.5 | |
| Python (0.75) sub-mm | 0.75 | 1.0 | 4 | 3.0 | sub-mm |
| Mt. Wilson (1.524) | 0.73 | 0.8 | 14 | 10.3 | |
| Mt. Wilson (2.54) | 0.70 | 1.0 | 16 | 10.9 | |
| FL Whipple (10) gamma ray | 0.66 | 2.5 | 22 | 14.3 | gamma-ray |
| UHawaii (2.2) | 0.61 | 7.4 | 95 | 58.2 | |
| CGRO (all instruments) | 0.59 | 29.1 | 251 | 148.0 | space |
| CBA East (0.66) | 0.59 | 0.5 | 6 | 3.3 | |
| Brigham Young Obs (BYO) (0.6) | 0.58 | 0.8 | 4 | 2.5 | |
| Stony Brook (0.36) | 0.57 | 0.2 | 1 | 0.7 | |
| IRAM (30) mm-wave | 0.56 | 10.7 | 84 | 46.9 | mm |
| ARGO balloon (1.2) | 0.56 | 1.0 | 9 | 5.0 | balloon |
| UKIRT (3.8) | 0.55 | 11.5 | 104 | 57.0 | |
| AT&T Bell Labs (7) radio | 0.52 | 0.4 | 6 | 3.2 | radio |
| NSO VTT solar | 0.50 | 4.0 | 24 | 12.0 | solar |
| Byurakan (1.0) Schmidt | 0.50 | 1.0 | 14 | 7.0 | |
| Kitt Peak NSO VT solar | 0.50 | 1.0 | 6 | 3.0 | solar |
| U. Missouri (0.35) | 0.50 | 1.0 | 4 | 2.0 | |
| Suhora (0.6) | 0.50 | 0.3 | 2 | 0.8 | |
| Kavalur (2.3) | 0.50 | 0.3 | 2 | 0.8 | |
| Foggy Bottom (0.4) | 0.50 | 0.2 | 1 | 0.7 | |
| Loiano (1.5) | 0.50 | 0.2 | 1 | 0.7 | |
| Kagoshima Space Center (0.6) | 0.50 | 0.2 | 1 | 0.7 | |
| Asiago Schmidt (0.40) | 0.50 | 0.3 | 1 | 0.3 | |
| Catania Schmidt (0.4) | 0.50 | 0.3 | 1 | 0.3 | |
| Other | 0.27 | 1200 | 12673 | 3424 | |
| Total (or average) | 0.30 | 1322 | 13632 | 4088 | |

**Table 11. Top telescopes, sorted by citations/paper.**
Again, small numbers of papers are often involved: but note also the highly cited papers.

| | citations/paper | papers (weighted) | pages | citations | |
|---|---|---|---|---|---|
| COBE (all instruments) | 17.00 | 1.0 | 12 | 17.0 | space |
| Mt. Wilson (1.524) | 13.67 | 0.8 | 14 | 10.3 | |
| DSN (34) radio | 13.00 | 0.2 | 3 | 2.6 | radio |
| Spartan 201 (solar) | 12.00 | 0.5 | 2 | 6.0 | solar |
| Mt. Wilson (2.54) | 11.42 | 1.0 | 16 | 10.9 | |
| CANGAROO (3.8) gamma ray | 11.00 | 0.5 | 2 | 5.5 | gamma-ray |
| Onsala (26) | 11.00 | 0.1 | 5 | 1.0 | |
| Hat Creek (26) radio | 10.00 | 0.5 | 5 | 5.0 | radio |
| Mt. Stromlo (1.3) | 9.50 | 2.0 | 16 | 19.0 | |
| Deep Space Network (DSN) radio (70) | 8.00 | 0.7 | 5 | 5.6 | radio |
| AT&T Bell Labs (7) radio | 8.00 | 0.4 | 6 | 3.2 | radio |
| UHawaii (2.2) | 7.84 | 7.4 | 95 | 58.2 | |
| Mees Solar coronagraph (0.25) | 7.60 | 0.8 | 6 | 6.3 | solar |
| UK Schmidt (1.2) | 7.47 | 3.0 | 45 | 22.1 | |
| Byurakan (1.0) Schmidt | 7.00 | 1.0 | 14 | 7.0 | |
| Owens Valley (40) radio | 6.93 | 0.9 | 13 | 6.1 | radio |
| CBA East (0.66) | 6.67 | 0.5 | 6 | 3.3 | |
| VU-TSU (0.4) | 6.60 | 0.6 | 21 | 3.7 | |
| Swedish solar VT (0.48) | 6.50 | 2.0 | 16 | 13.0 | |
| MSU (0.6) | 6.00 | 0.5 | 12 | 3.0 | |
| Cima Ekar (1.82) | 6.00 | 0.1 | 2 | 0.8 | |
| Torun (0.9) | 6.00 | 0.1 | 2 | 0.8 | |
| FL Whipple (10) gamma ray | 5.73 | 2.5 | 22 | 14.3 | gamma-ray |
| ASCA | 5.62 | 6.5 | 44 | 36.5 | space |
| ESO Dutch (0.9) | 5.40 | 0.8 | 6 | 4.5 | |
| Owens Valley mm array | 5.37 | 6.3 | 36 | 33.8 | mm |
| U Michigan radio (20) | 5.33 | 0.5 | 8 | 2.7 | radio |
| Siding Spring (2.3) | 5.24 | 5.5 | 63 | 29.0 | |
| CGRO (all instruments) | 5.09 | 29.1 | 251 | 148.0 | space |
| ARGO balloon (1.2) | 5.00 | 1.0 | 9 | 5.0 | balloon |
| UARS all instruments | 5.00 | 1.0 | 4 | 5.0 | space |
| ISAS (1.3) | 5.00 | 0.5 | 6 | 2.5 | |
| James Gregory (0.9) | 5.00 | 0.5 | 11 | 2.5 | |
| Yale-Columbia (0.66) | 5.00 | 0.2 | 3 | 1.0 | |
| Yale Southern (0.51) | 5.00 | 0.2 | 3 | 1.0 | |
| UKIRT (3.8) | 4.96 | 11.5 | 104 | 57.0 | |
| Keck I & II (10) | 4.84 | 22.0 | 125 | 106.3 | |
| Medicina radio (32) | 4.81 | 0.3 | 6 | 1.4 | radio |
| Noto radio (32) | 4.81 | 0.3 | 6 | 1.4 | radio |
| Westerbork radio array | 4.55 | 2.1 | 27 | 9.7 | radio |
| Other | 2.83 | 1207 | 12580 | 3416 | |
| Total (or average) | 3.09 | 1322 | 13632 | 4088 | |

**Table 12. Top telescopes, sorted by papers/collecting area.**
Radio telescopes and interferometers of all kinds have been excluded.

|  | papers/ area | pages/ area | citations/ area | total papers | |
|---|---|---|---|---|---|
| IUE (0.45) | 118.7 | 1426.1 | 394.2 | 33 | space |
| Mt. Hopkins APT (0.25) | 50.9 | 336.1 | 40.7 | 4 | |
| HST all instruments (2.4) | 28.5 | 288.7 | 107.1 | 163 | space |
| Burrell Schmidt (0.6) | 22.4 | 194.2 | 33.0 | 10 | |
| CTIO (0.9) | 19.6 | 234.2 | 39.0 | 25 | |
| Mees Solar coronagraph (0.25) | 17.0 | 122.2 | 129.0 | 2 | solar |
| KPNO (0.9) | 16.6 | 171.9 | 55.4 | 25 | |
| Kuiper Airborne (KAO) (0.91) | 13.8 | 129.9 | 39.2 | 10 | aircraft |
| CTIO Lowell (0.6) | 13.5 | 134.0 | 27.4 | 9 | |
| CBA West (0.36) | 13.1 | 49.1 | 13.1 | 2 | |
| CBA (0.32) | 12.4 | 126.4 | 49.7 | 4 | |
| Goethe-Link Roboscope (0.4) | 11.9 | 63.7 | 19.9 | 2 | |
| KPNO coude feed (0.9) | 11.8 | 200.5 | 27.7 | 12 | |
| CTIO Yale (1.02) | 11.3 | 130.7 | 19.9 | 20 | |
| Swedish solar VT (0.48) | 11.1 | 88.4 | 71.8 | 2 | |
| U. Missouri (0.35) | 10.4 | 41.6 | 20.8 | 1 | |
| Las Campanas Swope (1.02) | 10.3 | 116.8 | 31.5 | 13 | |
| Mt. Laguna (1.0) | 9.7 | 69.2 | 12.6 | 10 | |
| CTIO Curtis Schmidt (0.6) | 9.5 | 56.6 | 10.6 | 4 | |
| UHawaii (0.6) | 9.2 | 96.3 | 11.1 | 5 | |
| Palomar (1.52) | 8.9 | 116.2 | 24.4 | 27 | |
| Mentor (0.4) | 8.0 | 39.8 | 0.0 | 1 | |
| KPNO (1.3) | 7.7 | 87.5 | 19.6 | 21 | |
| KPNO (2.1) | 7.3 | 81.1 | 20.3 | 52 | |
| CTIO (1.5) | 7.3 | 85.4 | 21.5 | 27 | |
| Landis (0.2) | 7.0 | 49.3 | 10.5 | 3 | |
| Palomar (0.45) Schmidt | 6.3 | 150.9 | 12.6 | 1 | |
| SAAO (0.5) | 6.0 | 60.7 | 9.0 | 4 | |
| Rothney (0.4) | 6.0 | 63.7 | 2.0 | 2 | |
| Mt. Hopkins (0.4) | 5.7 | 36.2 | 2.6 | 4 | |
| Kiso Schmidt (1.05) | 5.7 | 61.9 | 11.5 | 7 | |
| Braeside (0.4) | 5.5 | 68.4 | 9.8 | 5 | |
| Mt. Laguna (0.6) | 5.3 | 37.1 | 3.5 | 2 | |
| Lick CAT (0.6) | 5.2 | 91.8 | 16.4 | 4 | |
| McGraw-Hill (1.3) | 5.2 | 40.8 | 8.5 | 13 | |
| Sproul (0.61) | 5.1 | 35.9 | 3.4 | 2 | |
| Toronto (0.61) | 5.1 | 24.0 | 5.1 | 2 | |
| Whipple (1.2) | 4.9 | 45.8 | 10.6 | 12 | |
| Dublin Obs (0.1) | 4.9 | 53.9 | 4.9 | 1 | |
| Steward (2.3) | 4.5 | 45.8 | 13.3 | 37 | |
| Other | | | | 1628 | |
| Total (or average) | | | | 2211 | |

**Table 13. Top ground-based, optical/IR telescopes, sorted by papers/collecting area**

| | total papers | f | papers (weighted) | pages | citations | pages/ paper | citations/ page | citations/ paper | papers/ area | pages/ area | citations/ area |
|---|---|---|---|---|---|---|---|---|---|---|---|
| Mt. Hopkins APT (0.25) | 4 | 0.63 | 2.5 | 17 | 2.0 | 6.60 | 0.12 | 0.80 | 50.9 | 336.1 | 40.7 |
| Burrell Schmidt (0.6) | 10 | 0.63 | 6.3 | 55 | 9.3 | 8.67 | 0.17 | 1.47 | 22.4 | 194.2 | 33.0 |
| CTIO (0.9) | 25 | 0.51 | 12.8 | 152 | 25.4 | 11.94 | 0.17 | 1.99 | 19.6 | 234.2 | 39.0 |
| Mees Solar coronagraph (0.25) | 2 | 0.42 | 0.8 | 6 | 6.3 | 7.20 | 1.06 | 7.60 | 17.0 | 122.2 | 129.0 s |
| KPNO (0.9) | 25 | 0.43 | 10.8 | 112 | 36.0 | 10.33 | 0.32 | 3.33 | 16.6 | 171.9 | 55.4 |
| CTIO Lowell (0.6) | 9 | 0.42 | 3.8 | 38 | 7.7 | 9.94 | 0.20 | 2.03 | 13.5 | 134.0 | 27.4 |
| CBA West (0.36) | 2 | 0.67 | 1.3 | 5 | 1.3 | 3.75 | 0.27 | 1.00 | 13.1 | 49.1 | 13.1 |
| CBA (0.32) | 4 | 0.25 | 1.0 | 10 | 4.0 | 10.17 | 0.39 | 4.00 | 12.4 | 126.4 | 49.7 |
| Goethe-Link Roboscope (0.4) | 2 | 0.75 | 1.5 | 8 | 2.5 | 5.33 | 0.31 | 1.67 | 11.9 | 63.7 | 19.9 |
| KPNO coude feed (0.9) | 12 | 0.64 | 7.7 | 130 | 18.0 | 16.93 | 0.14 | 2.34 | 11.8 | 200.5 | 27.7 |
| CTIO Yale (1.02) | 20 | 0.46 | 9.3 | 107 | 16.2 | 11.53 | 0.15 | 1.75 | 11.3 | 130.7 | 19.9 |
| Swedish solar VT (0.48) | 2 | 1.00 | 2.0 | 16 | 13.0 | 8.00 | 0.81 | 6.50 | 11.1 | 88.4 | 71.8 s |
| U. Missouri (0.35) | 1 | 1.00 | 1.0 | 4 | 2.0 | 4.00 | 0.50 | 2.00 | 10.4 | 41.6 | 20.8 |
| Las Campanas Swope (1.02) | 13 | 0.65 | 8.4 | 95 | 25.8 | 11.33 | 0.27 | 3.06 | 10.3 | 116.8 | 31.5 |
| Mt. Laguna (1.0) | 10 | 0.76 | 7.6 | 54 | 9.9 | 7.16 | 0.18 | 1.31 | 9.7 | 69.2 | 12.6 |
| CTIO Curtis Schmidt (0.6) | 4 | 0.68 | 2.7 | 16 | 3.0 | 5.93 | 0.19 | 1.11 | 9.5 | 56.6 | 10.6 |
| UHawaii (0.6) | 5 | 0.52 | 2.6 | 27 | 3.2 | 10.50 | 0.12 | 1.21 | 9.2 | 96.3 | 11.1 |
| Palomar (1.52) | 27 | 0.60 | 16.1 | 211 | 44.3 | 13.11 | 0.21 | 2.75 | 8.9 | 116.2 | 24.4 |
| Mentor (0.4) | 1 | 1.00 | 1.0 | 5 | 0.0 | 5.00 | 0.00 | 0.00 | 8.0 | 39.8 | 0.0 |
| KPNO (1.3) | 21 | 0.49 | 10.3 | 116 | 26.0 | 11.32 | 0.22 | 2.53 | 7.7 | 87.5 | 19.6 |
| KPNO (2.1) | 52 | 0.49 | 25.5 | 281 | 70.4 | 11.03 | 0.25 | 2.77 | 7.3 | 81.1 | 20.3 |
| CTIO (1.5) | 27 | 0.48 | 12.9 | 151 | 38.1 | 11.72 | 0.25 | 2.96 | 7.3 | 85.4 | 21.5 |
| Landis (0.2) | 3 | 0.07 | 0.2 | 2 | 0.3 | 7.07 | 0.21 | 1.51 | 7.0 | 49.3 | 10.5 |
| Palomar (0.45) Schmidt | 1 | 1.00 | 1.0 | 24 | 2.0 | 24.00 | 0.08 | 2.00 | 6.3 | 150.9 | 12.6 |
| SAAO (0.5) | 4 | 0.29 | 1.2 | 12 | 1.8 | 10.15 | 0.15 | 1.51 | 6.0 | 60.7 | 9.0 |
| Rothney (0.4) | 2 | 0.38 | 0.8 | 8 | 0.3 | 10.67 | 0.03 | 0.33 | 6.0 | 63.7 | 2.0 |
| Mt. Hopkins (0.4) | 4 | 0.18 | 0.7 | 5 | 0.3 | 6.33 | 0.07 | 0.46 | 5.7 | 36.2 | 2.6 |
| Kiso Schmidt (1.05) | 7 | 0.70 | 4.9 | 54 | 10.0 | 10.94 | 0.19 | 2.04 | 5.7 | 61.9 | 11.5 |
| Braeside (0.4) | 5 | 0.14 | 0.7 | 9 | 1.2 | 12.36 | 0.14 | 1.78 | 5.5 | 68.4 | 9.8 |
| Mt. Laguna (0.6) | 2 | 0.75 | 1.5 | 11 | 1.0 | 7.00 | 0.10 | 0.67 | 5.3 | 37.1 | 3.5 |
| Lick CAT (0.6) | 4 | 0.37 | 1.5 | 26 | 4.6 | 17.58 | 0.18 | 3.15 | 5.2 | 91.8 | 16.4 |
| McGraw-Hill (1.3) | 13 | 0.53 | 6.9 | 54 | 11.3 | 7.90 | 0.21 | 1.64 | 5.2 | 40.8 | 8.5 |
| Sproul (0.61) | 2 | 0.75 | 1.5 | 11 | 1.0 | 7.00 | 0.10 | 0.67 | 5.1 | 35.9 | 3.4 |
| Toronto (0.61) | 2 | 0.75 | 1.5 | 7 | 1.5 | 4.67 | 0.21 | 1.00 | 5.1 | 24.0 | 5.1 |
| Whipple (1.2) | 12 | 0.47 | 5.6 | 52 | 12.0 | 9.28 | 0.23 | 2.15 | 4.9 | 45.8 | 10.6 |
| Dublin Obs (0.1) | 1 | 0.04 | 0.0 | 0 | 0.0 | 11.00 | 0.09 | 1.00 | 4.9 | 53.9 | 4.9 |
| Steward (2.3) | 37 | 0.51 | 18.9 | 190 | 55.2 | 10.07 | 0.29 | 2.92 | 4.5 | 45.8 | 13.3 |
| McDonald (0.76) | 6 | 0.34 | 2.0 | 25 | 5.2 | 12.48 | 0.21 | 2.58 | 4.4 | 55.4 | 11.5 |
| Other | 892 | 0.44 | 476 | 5031 | 1544 | 10.57 | 0.31 | 3.25 | | | |
| Total (or average) | 1275 | 0.46 | 672 | 7135 | 2017 | 10.61 | 0.28 | 3.00 | | | |